# DESIGN AND CONSTRUCTION OF THE 3.2 MEV CATHODE ASSEMBLY FOR DARHT II*

C. Peters, S. Yu, S. Eylon, W. Ghiorso, E. Henestroza,
Lawrence Berkeley National Laboratory, Berkeley, CA 94720

## 1 INTRODUCTION

A 3.2 MeV injector has been designed and built for the second axis of the Dual Axis Radiographic Hydrodynamic Test (DARHT) facility at Los Alamos National Laboratory. Darht-II accelerator is a linac which produces a 2000 ampere electron pulse with a flattop width of at least 2-microseconds and emittance of less than 0.15 $\pi$ cm-rad normalized. The installation of the injector system is presently underway and commissioning is expected to begin in December 2000. The design and construction of the injector was completed by LBNL. The Marx Generator was designed and built under subcontract to LBNL. The installation of the insulating column will be occurring during August 2000.

The Darht-II injector High Voltage Column and Cathode Assembly are housed within a 4 meter diameter by 9 meter tall vacuum enclosure. The cathode assembly is supported on top of a vertically oriented insulating column. This arrangement is shown in figure 1.

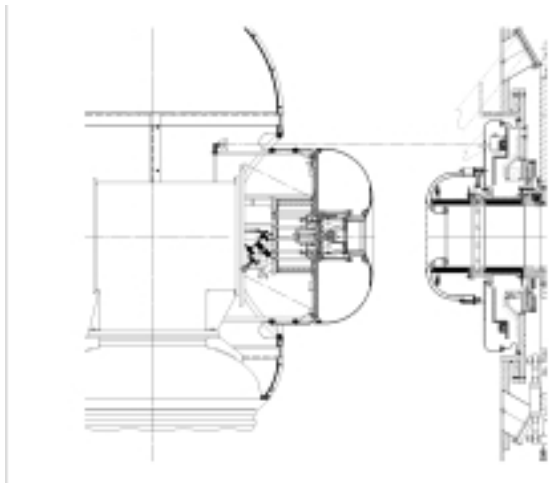

Figure 1 - Column and Cathode Assembly in Vacuum Enclosure

## 2 REQUIREMENTS

The basic requirement for the cathode assembly is to provide an electron source that meets the current and emittance requirements for the accelerator. This requires a source of sufficient size, current density, and current uniformity. The source must maintain accurate alignment to the accelerator. The shroud must have accurate geometry for beam electric focussing, must withstand high electric fields without breakdown, and must be accurately positioned with respect to the source. Additionally, the cathode assembly gas load must be consistent with the $5 \times 10^{-8}$ torr pressure requirement for the Injector. Provisions must be designed into the cathode assembly for easily maintaining the source, shroud, and associated utilities.

## 3 DESIGN

The arrangement of the source, shroud, and bucking solenoid is shown in Figure 2 and is referred to as the Front End Assembly. The source is a 6.5 inch diameter Osmium/ruthenium (M type) coated dispenser cathode with ohmic heater assembly and is similar to the sources used in klystrons. The heater is a single toroid of 0.100 inch diameter wire with center return. The design and temperature uniformity analysis was done under subcontract to LBNL. Predicted uniformity is ±10°C. The goal for current uniformity and density is <5% and 9.4 A/cm$^2$, respectively. These expectations are modest and are based upon ETA and more recent RTA experience. A cathode test stand at LANL will run a production source at full operating conditions starting in September.

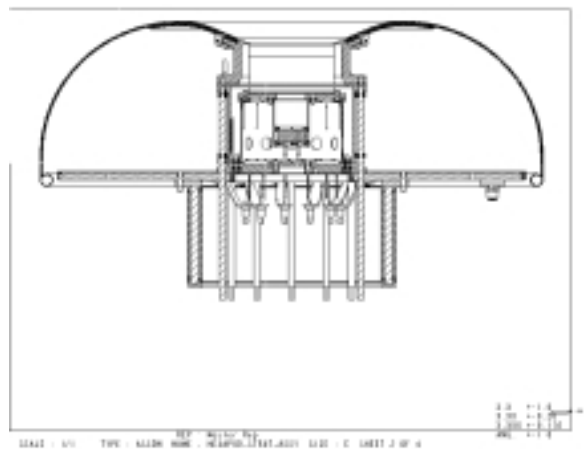

Figure 2 - Front End Assembly

*Work supported by the US Department of Energy under contract DE-AC03-76SF00098

The shroud is a 66° pierce design with an overall diameter of 1 meter. The peak field stress on the shroud surface is 120 kV/cm. The basic shape was formed by spinning, followed by CNC profile cut, and a final polish to achieve a 4-8 μin surface finish. The shroud is 317 stainless steel and has a copper disk (of 60 cm O.D.) brazed to the back (inside surface). The copper enables uniform heating of the high field area of the shroud to 200-250°C. This temperature is a compromise between the need to reduce barium plating on the shroud surface and at the same time reduce current emission from the shroud surface. The shroud is supported from a copper spool bolted to inside of the shroud. This spool also forms the inside edge of the piece surface. The temperature of this knife edge is low due its high thermal conductivity even though it has a 0.5mm gap to the 1100°C source. The 350 watts required to maintain the shroud at 200-250°C is captured from the sides of the source. The cylindrical shields around the source and the I.D. of the copper support, which has a UHV compatible black oxide coating, are designed in order to achieve the required radiant heat transfer.

The base of the source is supported by a stainless steel tube. The "hot" end of this tube is slit lengthwise to enable accommodation of the thermal expansion mismatch between the moly source base and the stainless steel tube.

The source will periodically be removed for maintenance or replacement. Source removal requires shroud removal. The shroud may also need to be removed for maintenance and for access to instrumentation and hardware contained within the cathode assembly. Both the source and the shroud are mounted on tapered flanges, which provide easy disassembly and precise and easy reassembly. The shroud "knife edge" is aligned to the edge of the source surface in the hot condition to ± 0.1 mm. Relative motion between these edges from cold to hot is predicted by the FEA model and taken into account in the cold geometry.

At the "cold" end of the source support tube are 12 pie shaped current viewing resistor assemblies, which enable determining the magnitude and uniformity of current flow to the source. These assemblies are thin sandwiches containing stainless steel strips and kapton insulating sheets.

The source and shroud support tubes are mounted to a 1 meter diameter base plate which is water cooled via drilled internal passages. To the upstream side of this plate is mounted the solenoid bucking coil.

The solenoid coil contains 12,000 amp-turns of 2mm solid square copper wire potted into a solid core. The 1500 watts of heat produced is cooled through water flowing over the I.D. and O.D. of the coil. The magnetic centerline of the coil was aligned to the source and shroud assembly at assembly.

## 4 CATHODE ALIGNMENT SYSTEM

The source, shroud, and solenoid elements are aligned to each other as described above. The requirement goal for this Front End Assembly is to be aligned to the accelerator centerline to within 0.2 mm in offset and 1 mrad in tilt. The long support path from the cathode assembly to the floor of the injector bay and back up to the accelerator centerline is about 60 feet. It is not possible to achieve the desired alignment accuracy for the cathode assembly through a simple passive support structure. Cathode position may be effected by building motion or by thermally induced structural changes. An active computer controlled alignment system has been built and tested.

The cathode assembly is supported by a hydraulically actuated hexapod. The position of the cathode with respect to the accelerator is measured using a system of lasers and detectors. Laser beams are reflected off mirrors mounted on the cathode assembly back onto position detectors mounted on the anode assembly. An algorithm computes cathode position errors based on the detector signals and computes hexapod strut length changes to correct the cathode position. Each hexapod location corresponds to a unique set of actuator strut lengths. Solenoid valves are then controlled to produce actuator length changes to achieve the computed target lengths. Actuator speed is about 0.001 inch/sec. Actuator lengths are known through linear potentiometers mounted on each actuator. After reaching target lengths the cathode position is re-read and the process continues until alignment of the cathode is within set tolerance. The final system works quite well; it converges and holds indicated position to ± 0.05mm. Presently, for sub-millimeter corrections, the control system takes 1-3 minutes to move into the aligned condition. Figure 3 shows a photo of the Hexapod assembly.

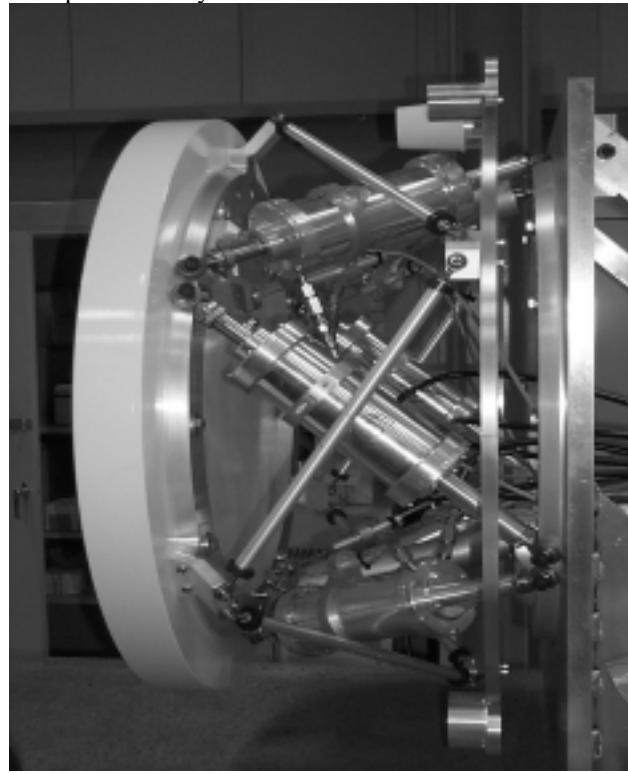

Figure 3 - Hexapod Assembly During Testing

There are 4 laser and detector pairs in the system. The TILT detector retro reflects the beam off a flat mirror and detects pitch and yaw errors. There are 2 XY-ROLL position detectors, which reflect off corner cubes. The reflected spot position moves 2x the cathode move. The Z detector incident and reflected laser beams form a 22° included angle and reflect off the same flat mirror as the retro reflector. The laser diodes reside in small enclosures mounted to the vacuum vessel anode area. Short fiber optic cables connect to the 4 beam output lenses. These components are shown in figure 4

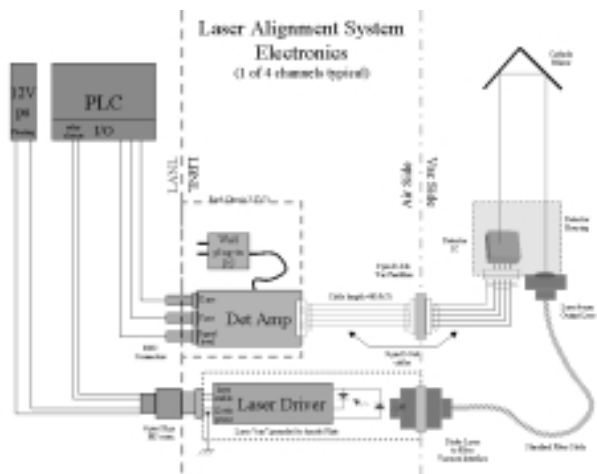

Figure 4 - Laser Alignment System Schematic

The actuators utilize welded 3 inch diameter bellows and have a ± 1.4 inch stroke which produces a ±1 inch X, Y, and Z travel for the cathode. The hydraulics work well in UHV since there are no elastomeric or sliding seals in the design. Hydraulic actuation was chosen over electrically actuated struts due to concerns over induced voltage and noise during injector operation. The working fluid is water at 100 psi. The actuators are single acting since loads are one directional and essentially constant. Auxiliary springs are used to supplement and help equalize actuator forces. Pressure is supplied through a small piston pump and accumulator combination. Compact commercial solenoid valves are used to control water in and out of the actuators. Pump and valves are housed within the marx dome inside the marx generator tank. Fluid lines and potentiometer signals are fed through the 5 meter current stalk which runs through the high voltage column into the vacuum space where the cathode assembly sits.

Initializing the alignment system at diode installation is fairly simple. The cathode is aligned to the anode at installation through use of a fixture, which physically mounts between the two assemblies and reads offset and tilt directly with dial indicators. The laser system is turned on and the PID outputs recorded. The control system compares PID signals to these reference values to determine misalignment.

During injector commissioning, the passive cathode position behavior will be observed and a best practice schedule for alignment corrections will be determined. Eventually, it may also turn out that there are reasons folks may want to run the injector with the cathode in positions other than the nominally aligned condition. The alignment system provides a way of accurately repeating cathode position in relation to the anode.